# Detecting network anomalies using machine learning and SNMP-MIB dataset with IP group

Abdelrahman Manna
*Princess Sumaya University of Technology*

manna.93@outlook.com

Mouhammd Alkasassbeh
*Princess Sumaya University of Technology*

m.alkasassbeh @psut.edu.jo

*Abstract*— SNMP-MIB is a widely used approach that uses machine learning to classify data and obtain results, but using SNMP-MIB huge dataset is not efficient and it is also time and resources consuming. In this paper, a REP Tree, J48(Decision Tree) and Random Forest classifiers were used to train a model that can detect the anomalies and predict the network attacks that my affect the Internet Protocol(IP) group. This trained model can be used in the devices that are used to detect the anomalies such as intrusion detection systems.

## I. INTRODUCTION

Nowadays, almost all of the world is connected to each other via internet and the number of internet users is increasing day by day, every user has at least 1-2 devices such as laptop or mobile phone.

As the number of users is increasing, the attacks on their devices are also increasing especially the attacks that affect the networks which is called "Network Attacks".

One of the most widely used and a well-known attack is the denial of service(DOS) that will be described in the coming section.

In this paper, a DOS attack is analyzed and the attacks on Internet Protocol(IP) group as a subset of SNMP-MIB groups that are described in [1].

### A. Network Attacks

Network attacks is a term that reflects and describes the attacks that may occur and affect the computer network in general, these attacks have big effects on the connected nodes as they might destroy the software that is installed on the connected node or prevent the connection from reaching to the node or from the node, this is also known as denial of service attack (DOS).
Denial of service attack can be described as an attack that affects the network to prevent the reach to the network resources such as server, this attack is considered as a dangerous attack because it prevents the legitimate users from reaching the resources whenever they need them, especially if the resource has sensitive and important information that need to be reached immediately.

### B. Simple Network Management Protocol (SNMP)

SNMP was found in late 1980s [2] is an application layer protocol that is used to control the functions of the network nodes(devices) in order to change their information or to change the devices' behaviours when needed, SNMP is supported by multiple devices such that routers, switches, servers and more and is included in the internet protocol(IP) package.
SNMP collects the data that needs to be managed and manages it using management information base(MIB) that describes the system configuration.

## II. RELATED WORK

One of the current hot topics in the network attacks is the DOS, researchers focus on anomaly detection for the anomalies that exploit the network and behave badly in order to prevent the legitimate nodes from connecting to the network or from reaching sensitive and important information.
In [3] showed in details the classification and technical analyses of network intrusion detection systems and the aspects that must be taken into consideration when using the Intrusion Detection Systems(IDS).
In [4] the authors showed one of the most commonly used techniques in detecting nodes that may affect the network and result in denial of service attack by using machine learning by training a model and give it a set of attacks with actual measures so the model can detect the anomalies or attack depending on the predefined datasets and results.
 The authors in [5] discussed machine learning technique for detecting the anomalies that uses the feature selection analysis that takes the top or most frequently used attacks and objects and classify them in a specific way that does not consume the network resources and does not exhaust them by enhancing the performance, but there is a probability of having false negative and false positive in the network.
In [6] authors showed ways for detecting the Distributed Denial Of Service(DDOS) attack which is more dangerous than the regular denial of service because the attacks come from different locations, the authors used a dataset and applied it on three classification techniques which are(Multilayer Perceptron (MLP), Naïve Bayes and Random Forest).
In [7] authors used predictive models and classifications for intrusion detection that use machine learning classifiers, they used Logistic Regression, Gaussian Naive Bayes, Support Vector Machine and Random Forest algorithms, their results showed that Random Forest gave the best results in classifying the traffic whether it is normal or not.

In [8] the author used MIB and Support Vector Machine(SVM) to achieve both high accuracy and fast detection and low false alarms.

III. RROPOSED MODEL

The proposed model is trained using Weka tool (V 3.8) that uses machine learning to achieve its results by training a model, in this paper three classifiers were used (Random Forest Algorithm, J48(Decision Tree), REP Tree Algorithm) to generate the results and check the accuracy of applying the IP group attacks on each classifier, noting that the results will be shown in the results section.

*A. SNMP-MIB Dataset*

In paper [4] the authors used a dataset that contains around 4998 records for 34 variables that are captured using MIB, paper [1] contains more information and description about the used dataset, in this paper the group that is used is taken from the dataset that is used in [4], and the used group is the internet protocol(IP) group, the attacks that are used are the attacks that may result in DOS attack which are(HTTP flood, UDP flood, ICMP-ECHO, TCP-SYN, Slowpost, Slowloris).

The MIB variables that are used in the IP group are described in Table I:

TABLE I. INTERNET PROTOCOL(IP) VARIABLE DESCRIPTION

| Variable Identifier | Variable Name | Variable description |
|---|---|---|
| Var1 | ipInReceives | The total number of input datagrams that are received from the interfaces, including the ones that are received in error. |
| Var2 | ipInDelivers | The total number of input datagrams that are delivered to the IP user protocols successfully(including ICMP). |
| Var3 | IpOutRequests | The total number of IP datagrams that are supplied to IP in requests for transmission, noting that this does not include the datagrams that are counted in ipForwDatagrams. |
| Var4 | ipOutDiscards | The number of output datagrams that do not have errors to prevent their transmission to their destination. |
| Var5 | ipInDiscards | The number of input datagrams that do not have errors to prevent their transmission to their destination. |
| Var6 | ipForwDatagrams | The number of input datagrams for which this entity was not their final destination. |
| Var7 | ipOutNoRoutes | The number of datagrams discarded because no route could be found to transmit them to their destination. |
| Var8 | ipInAddrErrors | The number of input datagrams discarded because the IP address in their header's destination field was not a valid address to be received at this entity. |

*B. Machine Learning Classifiers*

Machine learning classifiers are generated by an application in order to classify the attacks. The classifiers are mainly used build a model from classified objects and then use the same model to classify new ones that are not classified previously in the model in an accurate way as much as possible, the classifiers will be applied to classify the dataset that is used in this paper.

The used classifiers are considered as supervised learning algorithms that use labeled training data, the classifiers are described in details as follows:

- Random Forest Algorithm Classifier: Random Forest is a flexible and easy to use machine learning algorithm that gives great results most of the time. It is one of the most used algorithms because of its simplicity.
- J48 (Decision Tree) Classifier: Decision tree is also called information gain, a concept that measures the amount of information contained in a set of data. It gives the idea of importance of an attribute in a dataset.
- REP Tree Algorithm Classifier: REP Tree algorithm uses the regression tree logic then creates different multiple trees in different iterations, after generating the trees it chooses the best one from them and this is considered as the representative [1]

*C. Feature Selection*

The features are mainly used to reduce the computation time and to improve the performance of the model that is trained by minimizing the amount of data used, the feature selection strategy aims to remove the irrelevant fields to provide good results.

*Feature Selection Methods*

There are three methods for feature selection based on the evaluation criteria which are(Filter, Wrapper and Hybrid) that are defined by the authors in [9].
Filter methods are used as a step before the processing. Feature selection is independent of any machine learning algorithm. So, features are selected depending on their scores that are calculated from previous steps and statistics.
Wrapper methods are considered as selecting set of features as a search problem, this is done by combining different features together, and then give a score for them according to the accuracy of the model.
Hybrid methods are a combination of many feature selection methods such as filter and wrapper that are used together to achieve best results.

*Evaluation Metrics*

In this paper, a well-known evaluation criteria is used to measure the classifiers performance, such as F-Measure, accuracy, precision and recall.

The basic performance is indicated by the confusion matrix In Table II.

TABLE II. CONFUSION MATRIX FOR TWO CLASSES

| | | Predicted Class | |
|---|---|---|---|
| | | Positive | Negative |
| Actual Class | Positive | TP | FN |
| | Negative | FP | TN |

The weighted average for the accuracy of the classes are shown in (Table III), the weighted average for each of them is a result of all of the features that are used for IP group and calculated using WEKA tool and REP Tree classifier:

TABLE III. WEIGHTED ACCURACY RATE

| Accuracy Measure | TP Rate | FP Rate | Precision | Recall | F-Measure |
|---|---|---|---|---|---|
| Weighted Average | 1.0 | 0.0 | 1.0 | 1.0 | 1.0 |

True positive (TP) rate reflects the rate of the correct predictions of the positive traffic, while false positive (FP) rate reflects the rate of negative packets that are considered as positive traffic.

The true negative (TN) rate is the total number of negative traffic that is classified correctly as negative, while false negative (FN) rate shows the total number of positive traffic that is classified incorrectly as negative traffic.

$$Precision = \frac{TP}{TP + FP} \quad (1)$$

$$Recall = \frac{TP}{TP + FN} \quad (2)$$

$$F\text{-}Measure = 2\frac{Precision \cdot Recall}{Precision + Recall} \quad (3)$$

## IV. EXPERMINTAL RESULTS

The results are calculated by using WEKA tool 3.8 that uses machine learning to achieve its results by training a model, the specifications of the used hardware is Intel® CoreTM i7, 64-bits system with 8 GB RAM running on windows 10.

The experimental results are shown in this section, the results of the proposed model are generated from MIB dataset that is previously mentioned. The techniques of the classification are used to get results for the IP group separately from the main group. At the end, attribute selection techniques were used to enhance the accuracy of the proposed model by removing the irrelevant features and take the most relevant ones. This is used to show the impact of IP group on the classification of attacks.

The three classifiers' accuracy is shown in Table IV noting that REP Tree and Random Forest algorithms were more accurate that J48(Decision Tree).

TABLE IV. CLASSIFIERS' ACCURACY

| Classifier | Random Forest | J48 | REP Tree |
|---|---|---|---|
| Accuracy | 99.98% | 99.88% | 99.98% |

The F-Measure results for all of the IP group variables(V1,V2,V3,V4,V5,V6,V7,V8) are shown in Fig. 1, it can be noticed that in bruteforce attack the three used classifiers gave 1 which means that their accuracy for this attack is 100% while they are different in the other attacks.

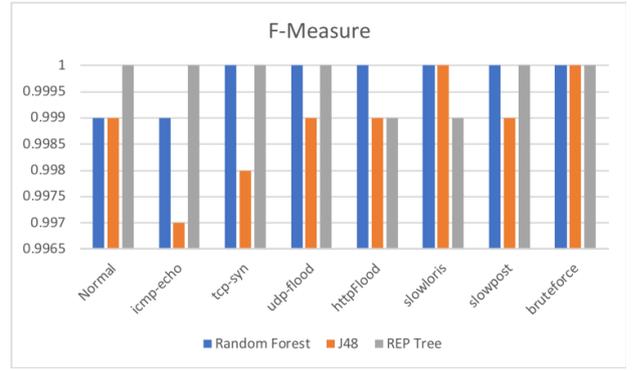

Fig. 1: F-Measure For All IP Group Variables

A top 5 and top 3 variables were selected to remove the most irrelevant variables from the IP group and were selected using InfoGainAttributeEval which evaluates the worth of an attribute by measuring the information gain with respect to the class, and by applying ranker method.

The results that are shown in Fig. 2 represent selecting the top 5 variables which are(V1,V4,V5,V6,V8):

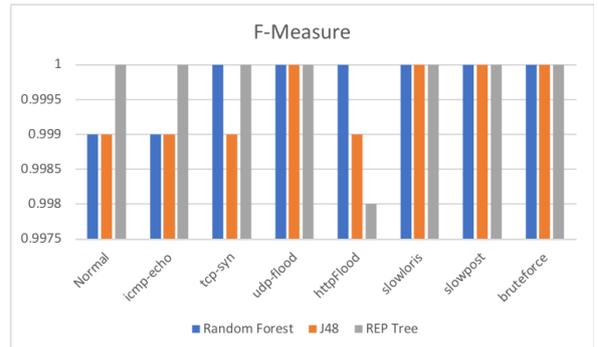

Fig. 2: F-Measure For Top 5 variables – InfoGain Attribute Evaluator

It can be noticed that the bruteforce percentage is still the same, but also the udp-flood, slowpost and slowloris and attacks gave 100% of accuracy.

The results that are shown in Fig. 3 represent selecting the top 3 variables which are(V1,V4,V5):

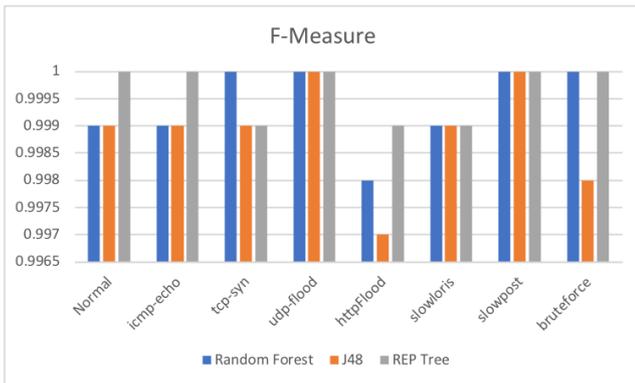

Fig. 3: F-Measure For Top 3 variables - InfoGain Attribute Evaluator

It can be noticed that the bruteforce attack accuracy were reduced in comparison with the above results, but udp-flood and slowpost attacks were still 100% accurate, this means that it is not necessary to have more accuracy when removing more irrelevant variables or reduce the training set size because in this experiment the top five variables gave more accuracy than selecting the top 3.

Another attribute evaluator is used to get the top 5 and top 3 variables which is ReliefFAttributeEval which evaluates the worth of an attribute by repeatedly sampling an instance and considering the value of the given attribute for the nearest instance of the same and different class.

The results that are shown in Fig. 4 represent selecting the top 5 variables which are(V1,V5,V6,V7,V8):

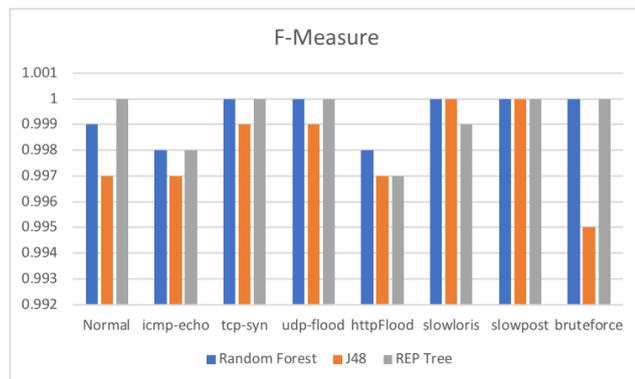

Fig. 4: F-Measure For Top 5 variables - ReliefF Attribute Evaluator

As shown in Fig. 4 the only attack which were 100% accurate was the slowpost.

The results that are shown in Fig. 5 represent selecting the top 3 variables which are(V6,V7,V8):

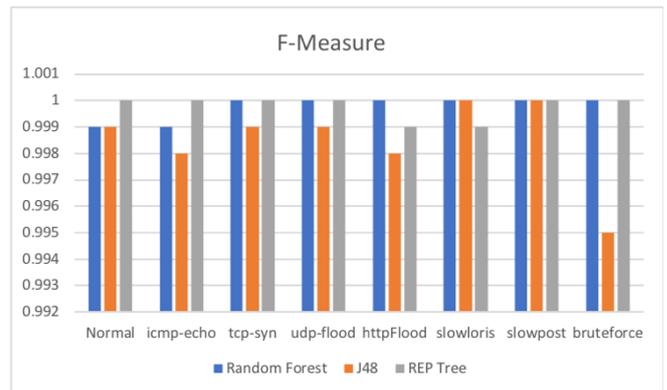

Fig. 5: F-Measure For Top 3 variables - ReliefF Attribute Evaluator

## V. CONCLUSION

In this paper, SNMP-MIB data were used to detect DOS attacks anomalies that may affect the network. Three machine learning algorithms were used to classify the data which are Random Forest, J48(Decision Tree) and REP Tree. Two Attribute evaluators were used to remove the irrelevant variables and get top 5 and top 3 variables, the two attribute evaluators are InfoGain and ReliefF. The classifiers and attributes were applied on the IP group and the results showed that applying the REP tree algorithm classifier gave the highest accuracy all of the times in all IP group, top 5 and top 3.